\documentclass[a4paper]{jpconf}
\usepackage{graphicx}
\begin{document}
\title{Gamma rays, electrons and positrons up to 3 TeV with the Fermi Gamma-ray Space Telescope}

\author{Philippe Bruel, on behalf of the Fermi-LAT collaboration}

\address{Laboratoire Leprince-Ringuet, Ecole polytechnique, CNRS/IN2P3, Palaiseau, France}

\ead{Philippe.Bruel@llr.in2p3.fr}

\begin{abstract}
The Fermi Gamma-ray Space Telescope (formerly known as Gamma-ray Large Area Space Telescope, GLAST) was successfully launched on June 11 2008. Its main instrument is the Large Area Telescope (LAT), which detects gamma rays from ~20 MeV to more than 300 GeV. It is a pair-conversion telescope with 16 identical towers (tracker and calorimeter), covered by an anti-coincidence detector to reject charged particles. The calorimeter is a hodoscopic array of CsI(Tl) crystals, arranged in 8 alternating orthogonal layers, with a total thickness of 8.6~radiation lengths. In this paper we will present the performance of the LAT, with special attention to the calorimeter, which provides a good energy measurement up to 3~TeV. We will also review some of its scientific results after 4 years of operation, focusing on measurements which extend up to very high energy, such as the spectrum of the extragalactic diffuse emission, the spectrum of cosmic electrons and the positron fraction.
\end{abstract}

\section{Introduction}
The main scientific goal of the Fermi Large Area Telescope~\cite{2009ApJ...697.1071A} is to study the gamma-ray sources from 20~MeV up to more than 300~GeV. There is a large variety of such gamma-ray sources, from galactic sources (pulsars, pulsar wind nebulae, supernova remnants, microquasars) to extragalactic sources (active galactic nuclei and gamma-ray bursts).  After 3 years of operation, the LAT had detected more than 1800 gamma-ray sources. On top of individual sources, we also seek to measure the galactic diffuse emission and the extragalactic diffuse emission.

The point source sensitivity is mainly the result of two characteristics of the telescope: its angular resolution and its total acceptance (the effective area times the field of view). The angular resolution drives the ability of finding a source on top of a relatively flat background. The total acceptance determines the number of gamma rays that can be detected from a given source. A large effective area is important, but a large field of view is also very important: it allows the telescope to observe a large portion of the sky at the same time, so that the total amount of time during which a given source is seen, is a large fraction of the total observational time of the telescope. Achieving the best angular resolution and the largest total acceptance were the main goals of the design of the LAT.

Regarding the energy measurement, the LAT was optimized to have the largest energy range: down to about 20 MeV and up to 300 GeV and more. In this energy range, gamma-ray sources generally have a smooth power-law energy spectrum with a photon index about -2. Thus measuring their energy spectra only requires a moderate energy resolution. But, since the spectra fall strongly with energy, any significant migration of reconstructed energy toward higher energies would easily distort the energy spectrum measurement. So the energy reconstruction must minimize the risks of overestimating the energy. Though most of the gamma-ray sources have a smooth energy spectrum, some, like pulsars and active galactic nuclei, have energy cutoffs or breaks that we want to measure as precisely as possible. We also want to reach the best sensitivity to gamma-ray lines that could result from dark matter annihilation. As a consequence, achieving the best energy resolution is also important for the LAT.

\section{From EGRET to the LAT}
In order to detect gamma rays above 10~MeV and measure their directions and energies, a space-based detector needs three sub-detectors: a tracker, in which gamma rays convert to pairs, to measure their directions, a calorimeter to measure their energies and an anti-coincidence detector (ACD) to reject charged particles. The tracker is made of some passive material in which the conversion occurs and some active material that tracks the $e^+e^-$ pair. The effective area depends on the fraction of converted gamma rays, which increases with the amount of converter. But below $\sim10$~GeV, the angular resolution is limited by multiple scattering, which also increases with the amount of converter. So a compromise must be found between increasing the effective area and degrading the angular resolution.

The predecessor of the LAT, EGRET~\cite{EGRET}, which operated between 1991 and 2000, had a tracker made of spark chamber modules interleaved with tantalum foils. Its volume was $80\times80\times45~\mathrm{cm}^3$ and the total on-axis depth was 0.54 radiation lengths, leading to an on-axis conversion efficiency of about 30\%. Between the tracker and the calorimeter, there was a time of flight coincidence system. So they were separated by $\sim80$~cm. As a consequence, the field of view of EGRET was limited to about $\sim25^\mathrm{o}$. The calorimeter of EGRET was a $77\times77\times20~\mathrm{cm}^3$ block of NaI, without segmentation. The total on-axis depth was about 8 radiation lengths. Its energy resolution was $\sim9$\% from 100~MeV to 10~GeV. The ACD of EGRET was a monolithic dome-shaped scintillator. At high energy, the electromagnetic showers in the calorimeter produce backsplash, which can produce some signal in the ACD, which mimics charged cosmic rays. So the maximum energy reached by EGRET was limited to about 20~GeV for two reasons: because the calorimeter was not segmented (preventing a good energy measurement) and because the ACD was not segmented (no ability to reject false vetos from backsplash).

\subsection{The Large Area Telescope}

The LAT was designed to overcome EGRET's limitations. It is composed of 16 identical towers arranged in a $4\times4$ array, with each tower made of a tracker module on top of a calorimeter module. The overall surface of the LAT is $150\times150~\mathrm{cm}^2$. A tracker module consists of 18 silicon (x,y) planes. The first 16 planes are interleaved with a tungsten foil, which thickness is 0.03 radiation lengths for the first 12 planes %(referred as front/thin section - optimized angular resolution at low energy)
and 0.18 radiation lengths for the following 4 planes% (referred as back/thick section - maximized conversion efficiency at the expense of a worse angular resolution)
. The total on-axis depth is 1.5 radiation lengths, leading to an on-axis conversion efficiency of about 70\%. The height of the tracker is $60~\mathrm{cm}$ and there are $5~\mathrm{cm}$ between the tracker and the calorimeter. So the overall aspect ratio of the LAT is much smaller than that of EGRET and its field of view is about $60^\mathrm{o}$. Each calorimeter module is an hodoscopic array of $8\times12$ CsI(Tl) crystals, arranged in 8 layers, each layer being $19.9~\mathrm{mm}$ thick. The total on-axis depth is 8.6 radiation lengths. Despite of its modest thickness, we will see in section~\ref{section:energyreconstruction} that the calorimeter segmentation allows us to measure gamma rays up to 3~TeV with good resolution. The ACD is segmented in 89 plastic scintillators tiles in order to minimize its sensitivity to backsplash. Thanks to its calorimeter segmentation and its ACD segmentation, the LAT is able to detect and measure gamma rays up 3~TeV, much higher than EGRET.

\subsection{The LAT calorimeter}
Each calorimeter module is made of 8 layers with 12 crystals per layer. Each crystal is a $326 \times 26.7 \times 19.9~\mathrm{mm}^3$ block of CsI(Tl). The orientation of the crystals depends on the layer: along {\it x} for the first layer, along {\it y} for the second, and so forth. This geometry allows for full imaging of gamma-ray showers. The crystal height is 1.07 radiation lengths and its width is 0.75 Moliere radii. This thin segmentation gives 2 coordinates of the energy deposition in a crystal. The third one, the longitudinal position along the crystal, is given by the asymmetry between the signals read out at each end.

In order to have a large energy range, each crystal end is readout by two PIN photodiodes ($25~\mathrm{mm}^2$ and $147~\mathrm{mm}^2$) with two gains per diode, leading to four ranges (2~MeV to 100~MeV, 2~MeV to 1~GeV, 30~MeV to 7~GeV, 30~MeV to 70~MeV). During normal data taking, the lowest non-saturated range is chosen. This very large dynamic range (2~MeV to 70~GeV) allows us to measure very low energy deposits (low energy gamma rays, minimum ionizing protons or ground muons) while avoiding crystal saturation for gamma rays up to $\sim1$~TeV. After 4 years of operation, there is only one dead diode out of 6144.

The in-flight calibration of the crystals is done in several steps. We use a charge-injection system to correct for electronics non-linearities. The lowest energy range is calibrated with the minimum ionizing cosmic proton peak at 11~MeV. The inter-range calibration is performed using non-interacting cosmic protons and heavy nuclei. The latter are also used to calibrate the asymmetry between the two ends, with the help of the tracker from which the reconstructed direction can be extrapolated to the crystal to give the longitudinal position reference. The observed crystal light yield attenuation due to radiation damage is about 1\% per year.

\begin{figure}[h]
\begin{minipage}{18pc}
\includegraphics[width=18pc]{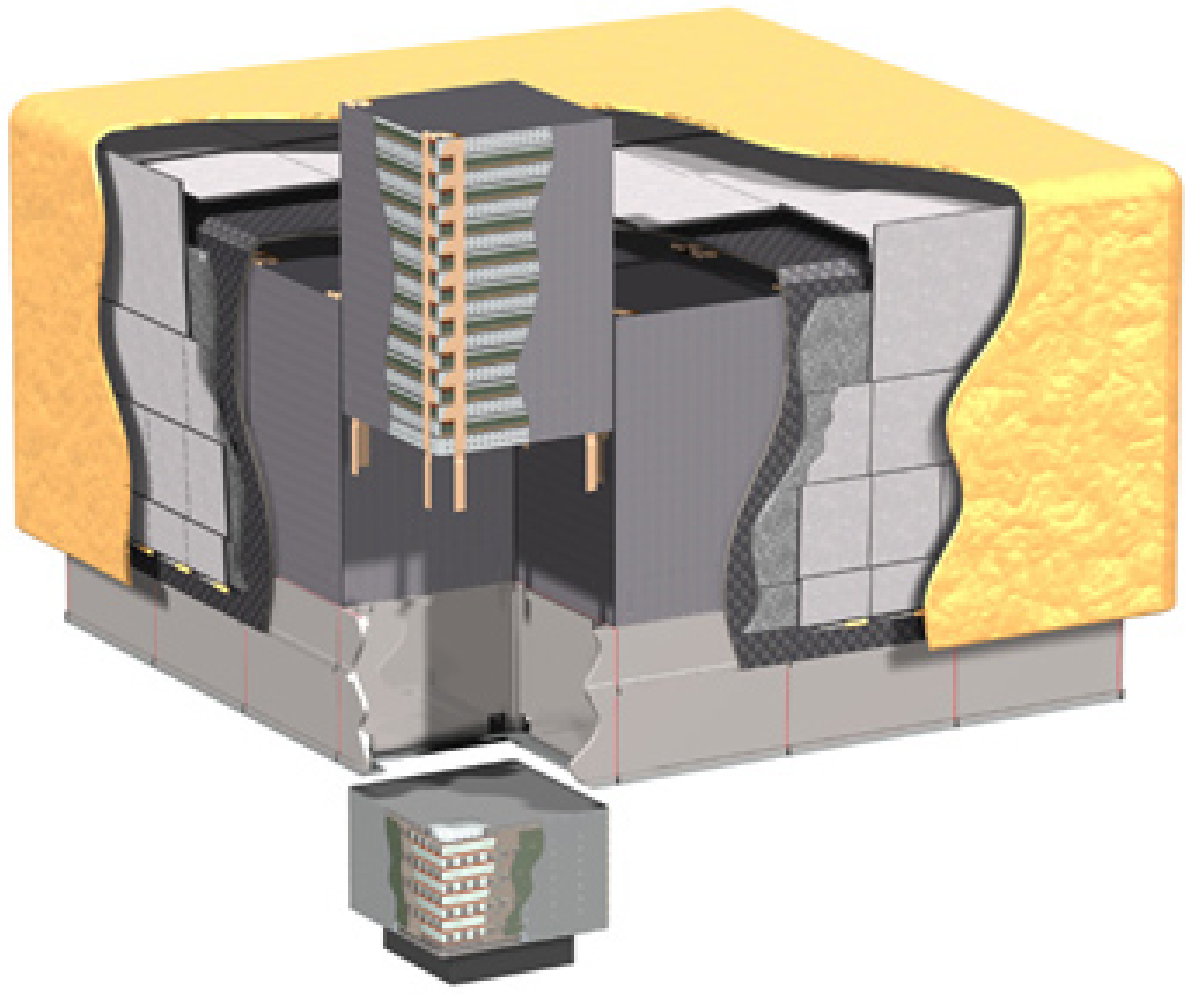}
\caption{\label{fig:lat}Schematic diagram of the Large Area Telescope.}
\end{minipage}\hspace{2pc}%
\begin{minipage}{18pc}
\includegraphics[width=18pc]{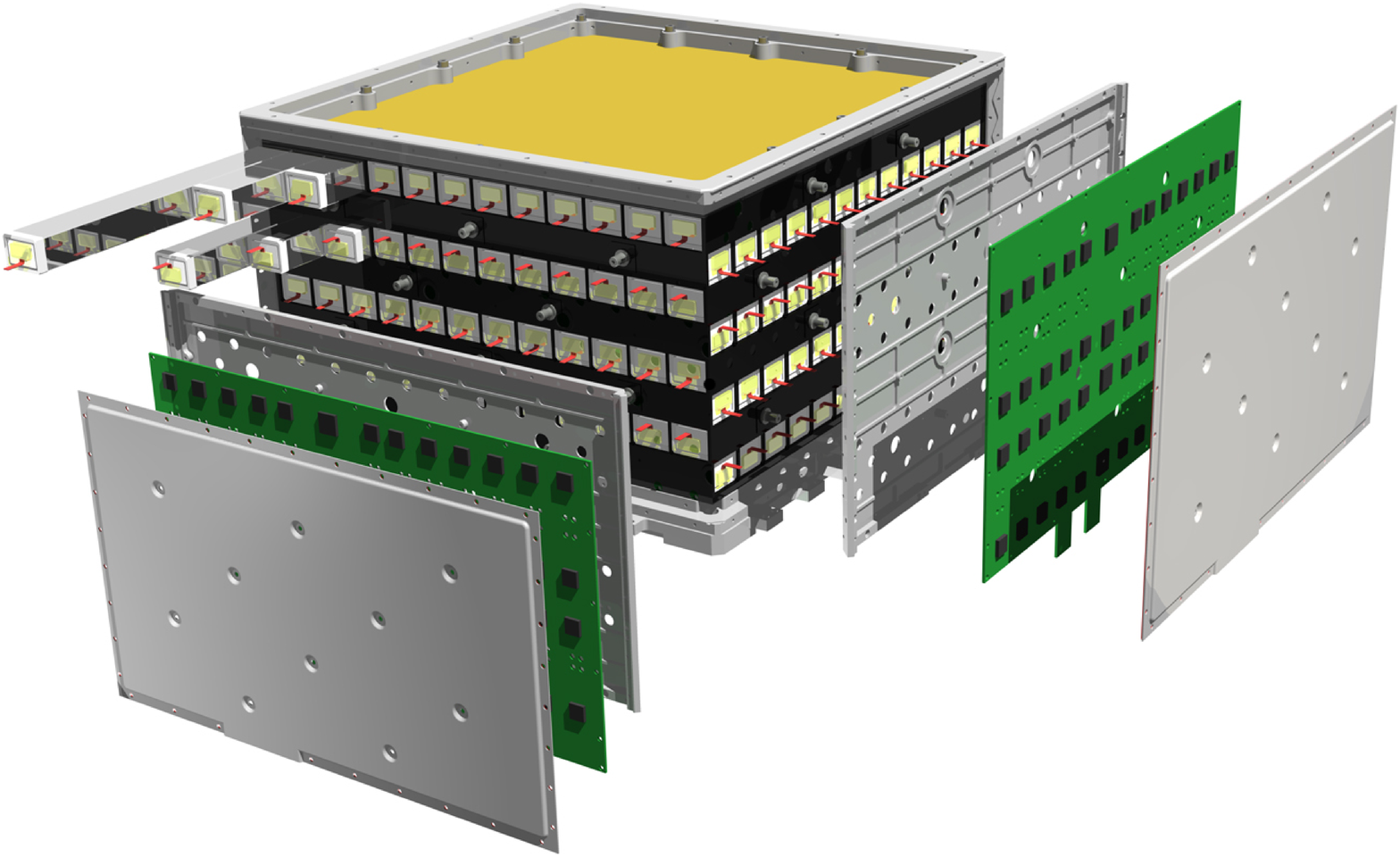}
\caption{\label{fig:latcal}Exploded view of a calorimeter module showing the hodoscopic $12\times8$ array of CsI detectors in the carbon fiber structure.}
\end{minipage} 
\end{figure}

\subsection{Event reconstruction and selection}
The goal of the LAT is to reconstruct and select good photon candidates. We use the tracker to find the direction of the event and the conversion point. The energy measurement relies on both the tracker and calorimeter. Very low energy gamma rays do not reach the calorimeter and their energy is determined from the number of hits in the tracker. For higher energies, we correct the energy deposited in the calorimeter for the various losses by using the position of the shower maximum or by performing a fit of the longitudinal profile of the shower, as fully described in section~\ref{section:energyreconstruction}.

The direction information is used to extrapolate the trajectory of the particle back to the ACD in order to reject charge particle background. This information, in addition to the topology of the shower in the tracker and in the calorimeter, is used for the selection of the gamma-ray candidates. The reconstruction and selection are based on classification trees and have been optimized using a detailed GEANT4 simulation. The performance of the instrument has been carefully checked in beam tests and on orbit~\cite{latperformance}. The on-axis effective area is $0.25~\mathrm{m}^2$ at 100~MeV, $0.7~\mathrm{m}^2$ at 1~GeV and $0.8~\mathrm{m}^2$ at 100~GeV.

The rate of cosmic-ray background passing through the LAT can be as high as 10~kHz. Since the characteristic signal shaping times of the tracker, calorimeter and ACD are respectivelly $10, 3.5$ and $4~\mu\mathrm{s}$, such a rate implies that the probability that particles pass through the LAT a few $\mu\mathrm{s}$ before a gamma ray triggers the detector, can be higher than 10\%. These out-of-time particles can leave sizable signals in all the three subdetectors. These spurious signals can cause serious errors in the event analysis and thus must be cleaned by the event reconstruction. In the calorimeter this is done by running a clustering algorithm and finding the clusters related to the real event.

\section{Energy reconstruction from 3~GeV up to 3~TeV}
\label{section:energyreconstruction}

The LAT phase space is very large: energies from 20~MeV up to more than 300~GeV and angles up to 60~deg with respect to the vertical axis of the detector. Because of the geometry of the instrument, the fraction of energy deposited in the calorimeter varies very much throughout this phase space. For normally incident photons, the fraction of energy deposited in the calorimeter is maximum at 1~GeV and is about 80\%. Above 1~GeV, leakage from the calorimeter starts to be the main cause of energy loss. Another cause of energy loss is the 4~cm gaps between the towers. Here we describe an energy reconstruction method that works for photons above $\sim3$~GeV, when the energy deposited in the tracker is small compared to that in the calorimeter, and where the leakage from the calorimeter can be large.

The longitudinal segmentation of the calorimeter allows a fit of the shower longitudinal profile, which is a good way to correct for energy loss when the shower is not fully contained in the calorimeter. In order to be able to perform such a fit, we must predict the energy deposited in each layer for any longitudinal profile. At normal incidence, the layers are perpendicular to the shower axis and the prediction is obvious. But at large incidence angle, the energy contained in a slice of the shower at a given age is shared among several layers or part of it escapes the calorimeter. As a consequence, for a given shower, we have to model accurately its development through the layers of the calorimeter, taking into account the gaps between towers.

The crystal electronic signals saturate at about 70~GeV. This saturation, which occurs for photons above 1~TeV, must be taken into account by the shower fit. When a crystal is saturated, we do not know how much energy has been deposited in it, so it cannot be used in the shower fit. Since the saturated crystals lie along the core of the shower, the only information we have comes from the transverse tails of the shower, which contain less than 10\% of the energy. That means that, in order to provide a useful energy resolution above 1~TeV, we have to be able to model very accurately the development of the shower through the calorimeter at the crystal level, and paying close attention to those crystals that lie at more than one Moliere radius from the shower axis.

As a consequence, the shower longitudinal profile fit is based on the modelling of the longitudinal and transverse profiles of electromagnetic showers and on the modelling of the development of the showers through the LAT calorimeter.

\subsection{Shower longitudinal profile parameterization}
\label{section:showerlongitudinalprofile}

The average longitudinal shower profile can be described by a gamma function~\cite{Longo:1975wb}:
\begin{equation}
\label{eq:gammaprofile}
\left< \frac{dE(t)}{dt} \right> = P_{(E,\alpha,\beta)}(t) = E \times \frac{(\beta t)^{\alpha-1} \beta e^{-\beta t}} {\Gamma(\alpha)}
\end{equation}
where $t$ is the longitudinal shower depth in units of radiation length, $\alpha$ the shape parameter and $\beta$ the scaling parameter. The position of the maximum of the profile is $T = (\alpha -1)\beta$. As shown in~\cite{Grindhammer:1993kw}, the same function can be used to describe individual profiles. In order to model the longitudinal profile, we have to determine how the distributions of $\alpha$ and $\beta$ vary with energy.

In order to do so, simulations of photons in a very large CsI calorimeter with {\tt GEANT4} at various energies have been performed. The calorimeter is segmented in 400 layers of 1.85~mm = 0.1 radiation length. Its total depth is 40 radiation lengths so photons are fully contained. For each event, the conversion point is used to define the shower starting point $t = 0$. In order to smooth the longitudinal profile, 0.5 radiation length bins were used to construct the histogram of the longitudinal profile, which is fitted with equation~(\ref{eq:gammaprofile}).

\begin{figure}[h]
\begin{minipage}{18pc}
\includegraphics[width=18pc]{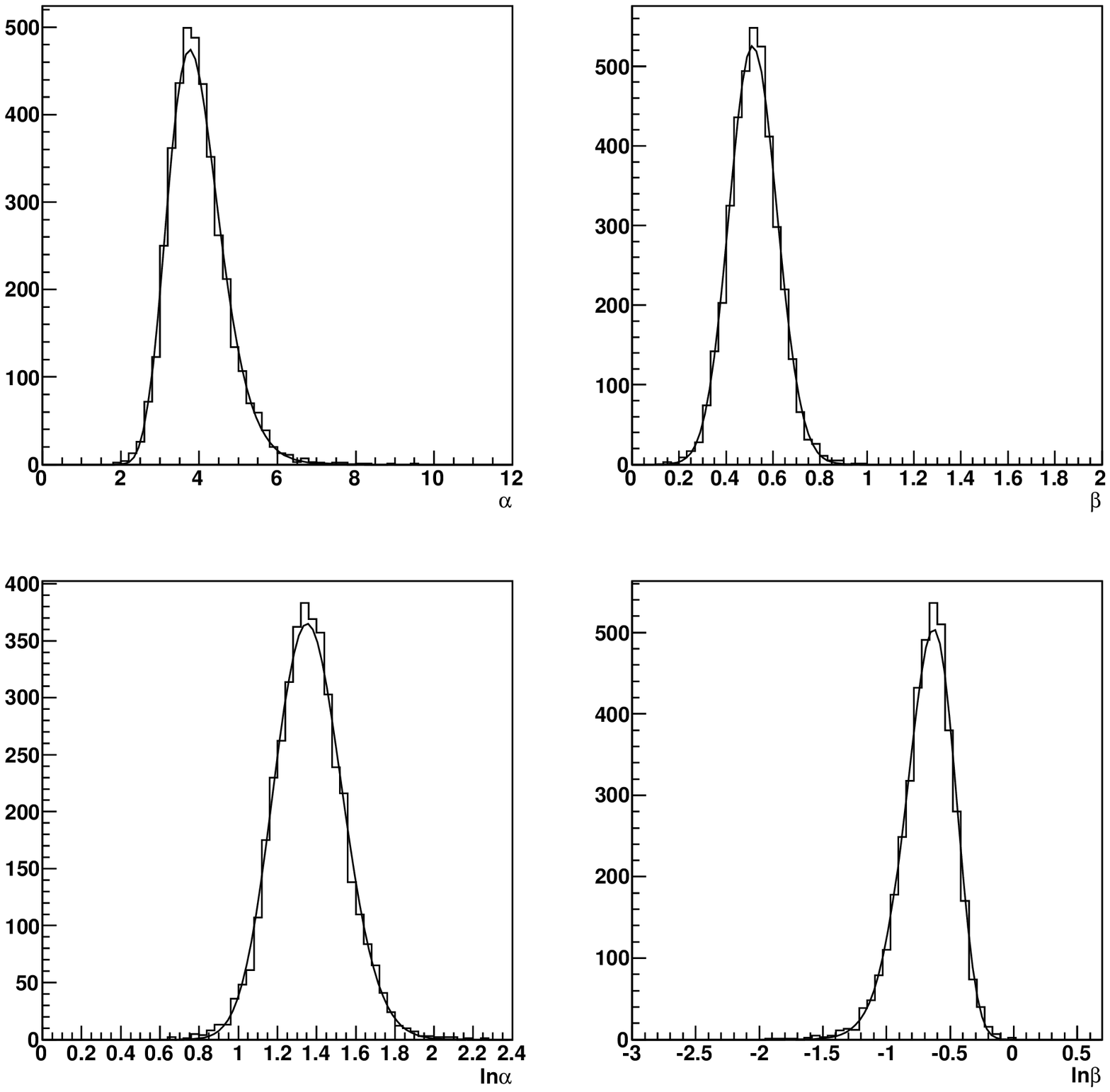}
\caption{\label{fig:alphabeta}Distributions of $\alpha$ and $\beta$ and their logarithms for 10~GeV photons.}
\end{minipage}\hspace{2pc}%
\begin{minipage}{18pc}
\includegraphics[width=18pc]{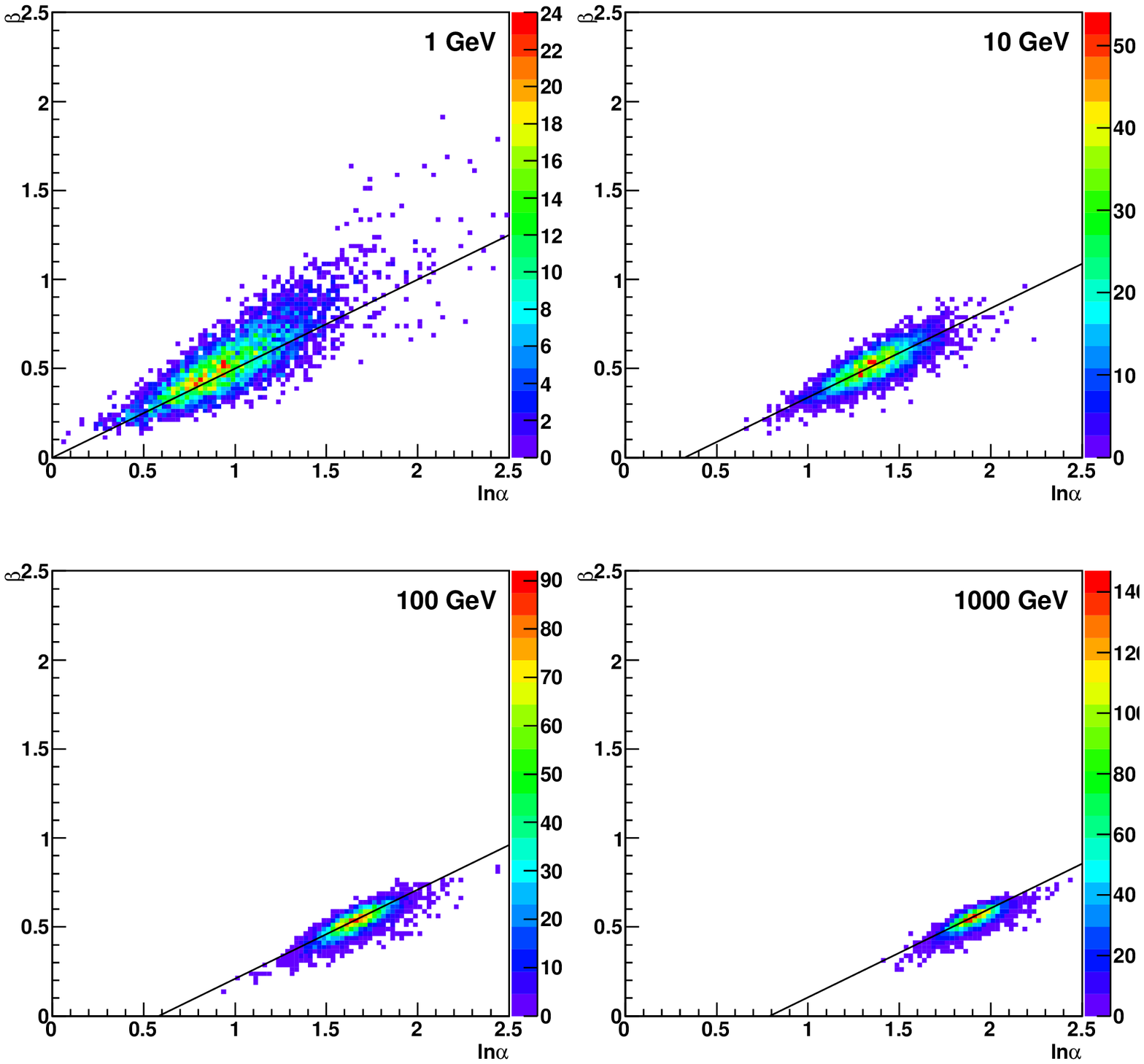}
\caption{\label{fig:lnalphabetacorrelation}Correlation between $\ln{\alpha}$ and $\beta$ for 1,10,100 and 1000~GeV photons.}
\end{minipage} 
\end{figure}

Fig~\ref{fig:alphabeta} shows the distributions of $\alpha$ and $\beta$ and their logarithm for 10~GeV photons. The distribution of $\ln{\alpha}$ is more gaussian than $\alpha$, whereas the distribution of $\beta$ is more gaussian than $\ln{\beta}$. That is why we decide to use $\ln{\alpha}$ and $\beta$.

\begin{figure}[h]
\begin{minipage}{18pc}
\includegraphics[width=18pc]{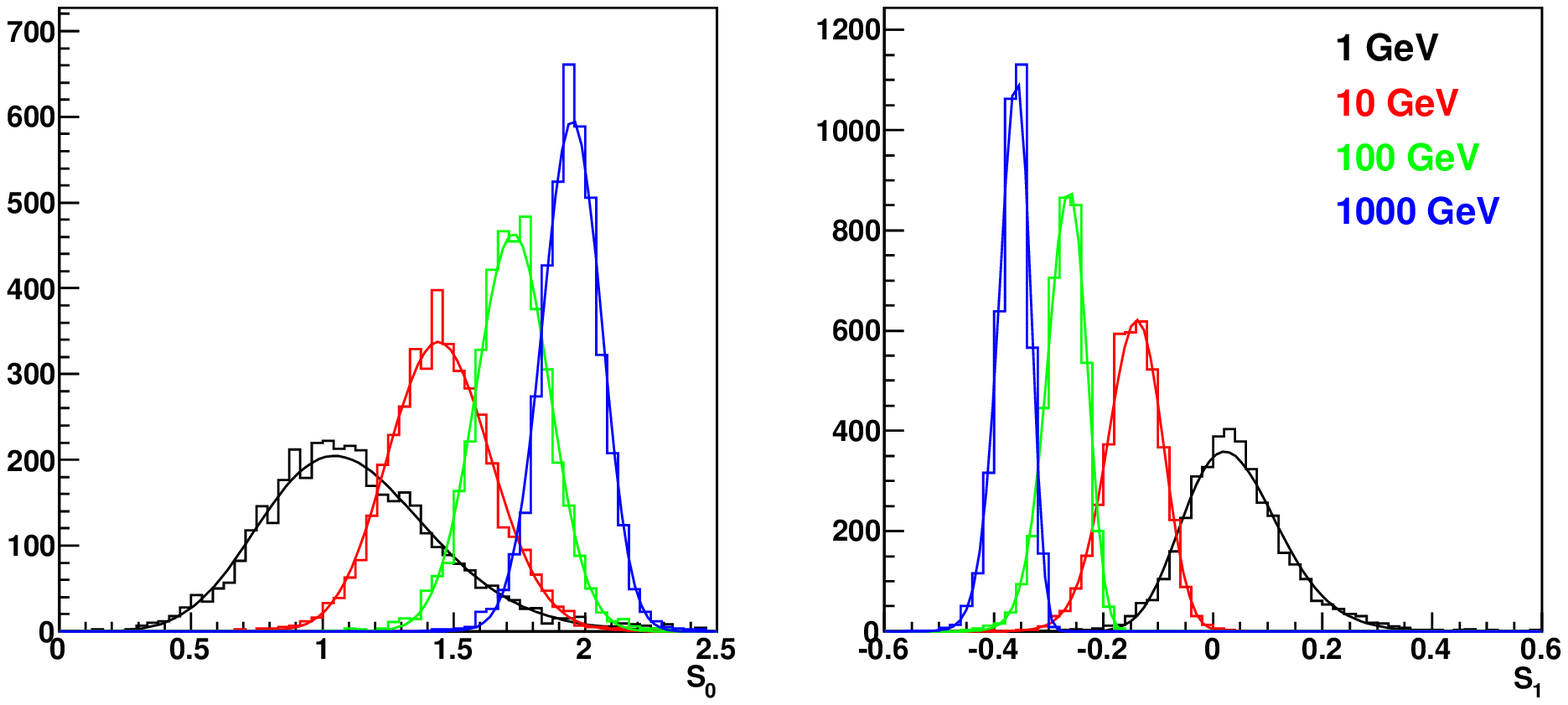}
\caption{\label{fig:ab}Distributions of the shower parameters $S_{0}$ and $S_{1}$ for 1, 10, 100 and 1000~GeV photons.}
\end{minipage}\hspace{2pc}%
\begin{minipage}{18pc}
\includegraphics[width=18pc]{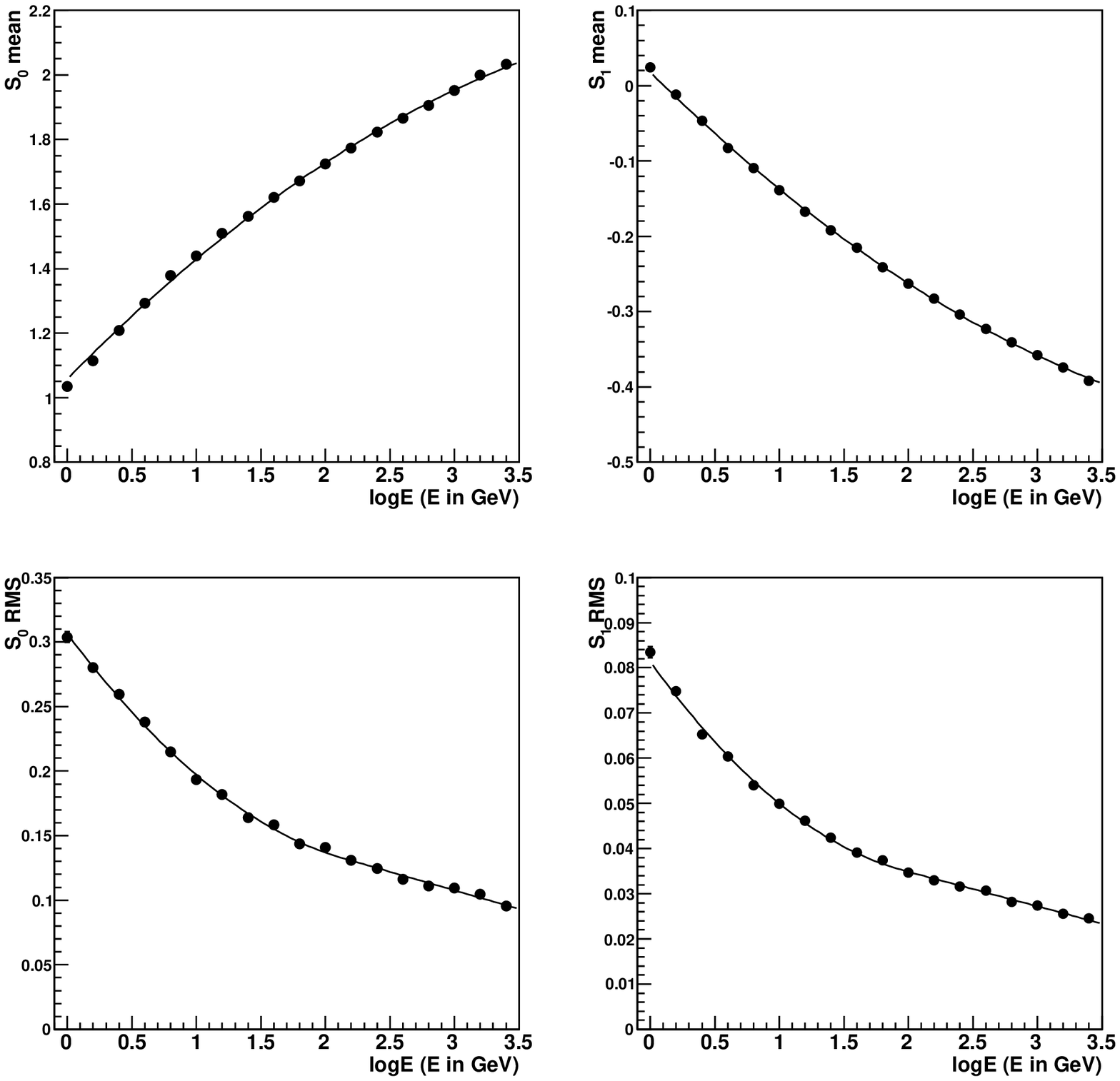}
\caption{\label{fig:s0s1param}Parameterization of the mean and RMS of $S_{0}$ and $S_{1}$ as function of logE.}
\end{minipage} 
\end{figure}

Fig~\ref{fig:lnalphabetacorrelation} shows the correlation of these two variables for various energies. The correlation is independent of energy with a slope of $\tan{\theta_{c}} = 0.5$. As it is preferable to use uncorrelated variables, we use the following variables: 
$$
\begin{array}{ccl}
  S_{0} & = & \ln{\alpha}\cos{\theta_{c}}+\beta\sin{\theta_{c}} \\
  S_{1} & = & -\ln{\alpha}\sin{\theta_{c}}+\beta\cos{\theta_{c}}
\end{array}
$$
Fig~\ref{fig:ab} shows the distributions of $S_{0}$ and $S_{1}$ for various photon energies. The last step of the description of the shower parameters $S_{0}$ and $S_{1}$ is the parameterization of the variation of their mean and RMS with logE, as shown in Fig~\ref{fig:s0s1param}. The dependence of the mean can be simply parameterized with a quadratic polynomial. The dependence of the RMS is quadratic below $\sim30$~GeV and linear above.

In order to determine the error of the modeling of the shower profile, that is to say the error that will be used in the calculation of the $\chi^2$ during the fit, we use the residuals of the shower profile fit as a function of the shower depth $t$. The RMS of these residuals are almost flat with a broad maximum at $0.5<t/T<1.5$. As a consequence we assume that the error does not depend on the shower depth and we use the same modeling error for all layers. We define $\epsilon_{T}$ as the RMS of the residuals at shower maximum divided by the energy deposited at shower maximum. Again $\epsilon_{T}$ varies with energy. Its variation can be parameterized by $\epsilon_{T}(E) = 0.17e^{-\log{E}/1.38}$, with E in GeV. During the fit, the error used in the $\chi^2$ calculation is $\delta e = \epsilon_{T}(E) \times \mathrm{max}(E_i)$, where $E_i$ is the energy deposited in layer $i$.

\subsection{Shower transverse profile parameterization}
\label{section:showertransverseprofile}

The average transverse energy profile can be described by two components, a core and a tail. As in~\cite{Grindhammer:1993kw}, we use the following function:
$$ f(t/T,r) = \frac{1}{dE(t)} \frac{dE(t,r)}{dr}  = p \frac{2rR_C^2}{(r^2+R_C^2)^2} + (1-p) \frac{2rR_T^2}{(r^2+R_T^2)^2} $$
where $R_C$ and $R_T$ are the medians of the core and the tail components (in units of Moliere radius), and $p$ is the relative weight of the core component ($0<p<1$). This transverse profile depends on the shower depth $t/T$. In order to study the dependence of $R_C$, $R_T$ and $p$ on $t/T$, simulations were performed with {\tt GEANT4}. In this case, the CsI calorimeter was divided longitudinally in slices of 0.1 radiation length and radially in rings with a width of 0.1 Moliere radius, which is 3.5~cm for CsI. For each event, we perform the longitudinal profile fit in order to determine $T$, the position of the shower maximum.

We compute the average fraction of energy contained in a cylinder as a function of the cylinder radius for various intervals of the shower depth $t/T$. Fig.~\ref{fig:transverseprofile}(left) shows some examples of these cumulative transverse profiles for 100~GeV photons. The fit of these profiles allows the determination of $R_C$, $R_T$ and $p$. Fig.~\ref{fig:transverseprofile}(right) shows how these parameters vary with $t/T$. The variation of the average transverse profile with energy was found to be very weak, so we neglect it and use the 100~GeV profile.

\subsection{Constraining the longitudinal fluctuations during the fit}
\label{section:profilefitprinciple}

The fit of the shower longitudinal profile works well when the shower maximum is well contained in the calorimeter. Since the LAT calorimeter is composed of 8 layers, this is the case when there is significantly less energy in the last layer than in the seventh layer. For incidence angles smaller than 25~deg, the fraction of photons for which the shower maximum is not well contained is $\sim25$\% for photons at 100~GeV. The situation above 1~TeV is even worse since almost no photons within 45~deg of the detector axis have a well contained shower maximum.

\begin{figure}[h]
\begin{minipage}{18pc}
\includegraphics[width=18pc]{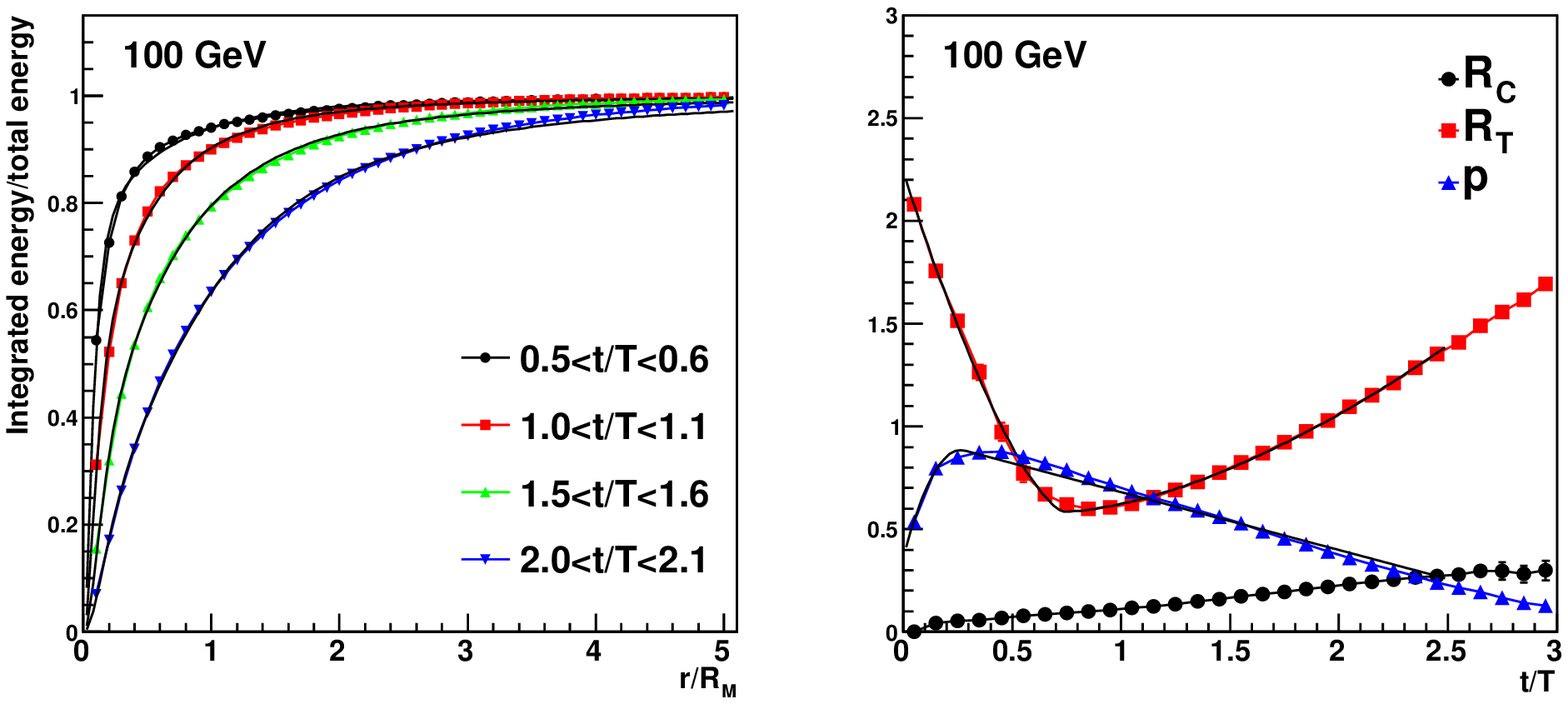}
\caption{\label{fig:transverseprofile}Left: average transverse profiles at 100~GeV. Right: variation of the parameters of the transverse profile with the position along the shower at 100~GeV.}
\end{minipage}\hspace{2pc}%
\begin{minipage}{18pc}
\includegraphics[width=18pc]{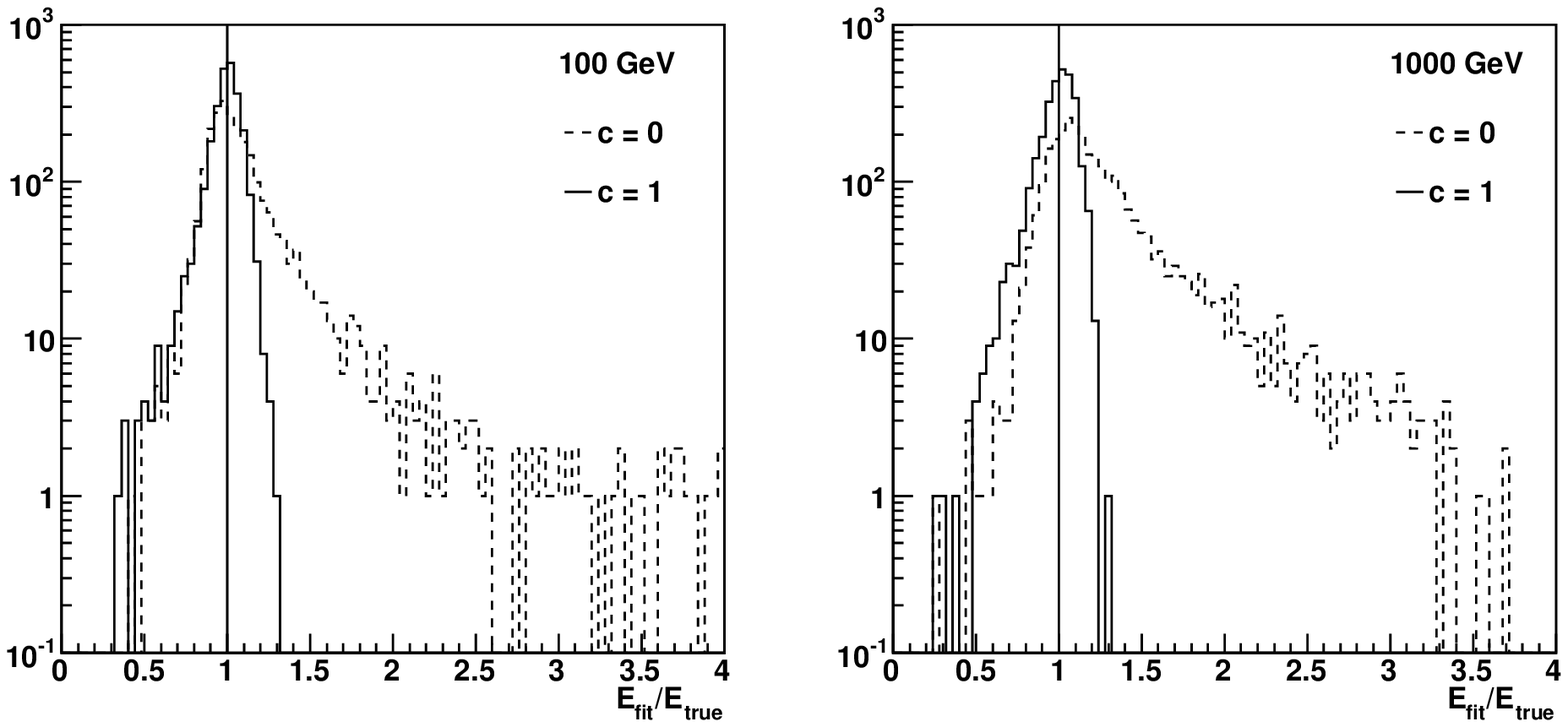}
\caption{\label{fig:constraintweight2}Comparison of the distributions of $E_\mathrm{fit}/E_\mathrm{true}$ with and without the parameter constraining term at 100~GeV (left) and 1~TeV (right).}
\end{minipage} 
\end{figure}

In order to be able to reconstruct the energy of these photons, we have to use more information than the deposited energies in the layers. The idea is to use the knowledge on the fit parameters $S_{0}$ and $S_{1}$ (especially the dependence of their mean and RMS with E) by adding to the $\chi^2$ a term that constrains the parameters to be close to their expected values. Since $S_{0}$ and $S_{1}$ are gaussian and uncorrelated, we use $s^2_{0}+s^2_{1}$, where $s_i(E) = (S_i-\mu_{S_i}(E))/\sigma_{S_i}(E)$, i.e. the distance of $S_i$ to its mean value and divided by its RMS. The shower fit $\chi^2$ becomes:
$$ \chi^2(S_{0},S_{1},E) = \sum_{i=1}^8 \frac{\left(e_{m,i}-e_{p,i}(E) \right)^2}{\delta e^2(E)} + c\left( s^2_{0}(E)+s^2_{1}(E) \right) $$
where $e_{m,i}$ and $e_{p,i}$ are respectively the measured and predicted energies in layer $i$. The parameter constraint term is multiplied by a factor $c$. We varied $c$ and found that $c=1$ was optimal. Fig.~\ref{fig:constraintweight2} compares the distributions of $E_\mathrm{fit}/E_\mathrm{true}$ with $c=0$ and $c=1$, for 100 and 1000~GeV photons at normal incidence. When $c=0$, the energy can be overestimated by a factor greater than 2, whereas setting $c=1$ allows us to remove entirely the high energy tail, even when the shower maximum is not well contained.

\subsection{The development of showers through the LAT calorimeter}
\label{section:devthroughcal}

For a given shower longitudinal profile $P_{(E,\alpha,\beta)}$, the predicted energy deposited in layer $i$ is:
$$ e_{p,i} = \int_0^\infty f_i(t)P_{(E,\alpha,\beta)}(t) \mathrm dt $$
where $f_i(t)$ is the fraction of energy deposited in layer $i$ by the shower slice between $t$ and $t+ \mathrm dt$. For normal incidence photons, $f_i(t)$ only depends on the position of the shower slice $t$. But when the incidence angle is greater than 0, the layers are no longer perpendicular to the axis of the shower and so the transverse profile must be taken into account. Since the transverse profile is a function of $t/T$, the energy fraction $f_i(t)$ also depends on $T$.

The computation of $f_i(t)$ depends on the photon trajectory. If the photon converted in the tracker, we use the results of the tracking algorithm to get the direction and the conversion point, corresponding to the starting point of the shower. We then compute the number of radiation lengths crossed by the shower between the conversion point and the extrapolated entry point of the photon in the calorimeter. If no track has been found, we use the results of the moments analysis of the shower in the calorimeter to get the direction and we assume that the conversion point corresponds to the entry point of the photon in the calorimeter.

\begin{figure}[h]
\begin{minipage}{16pc}
\includegraphics[width=16pc]{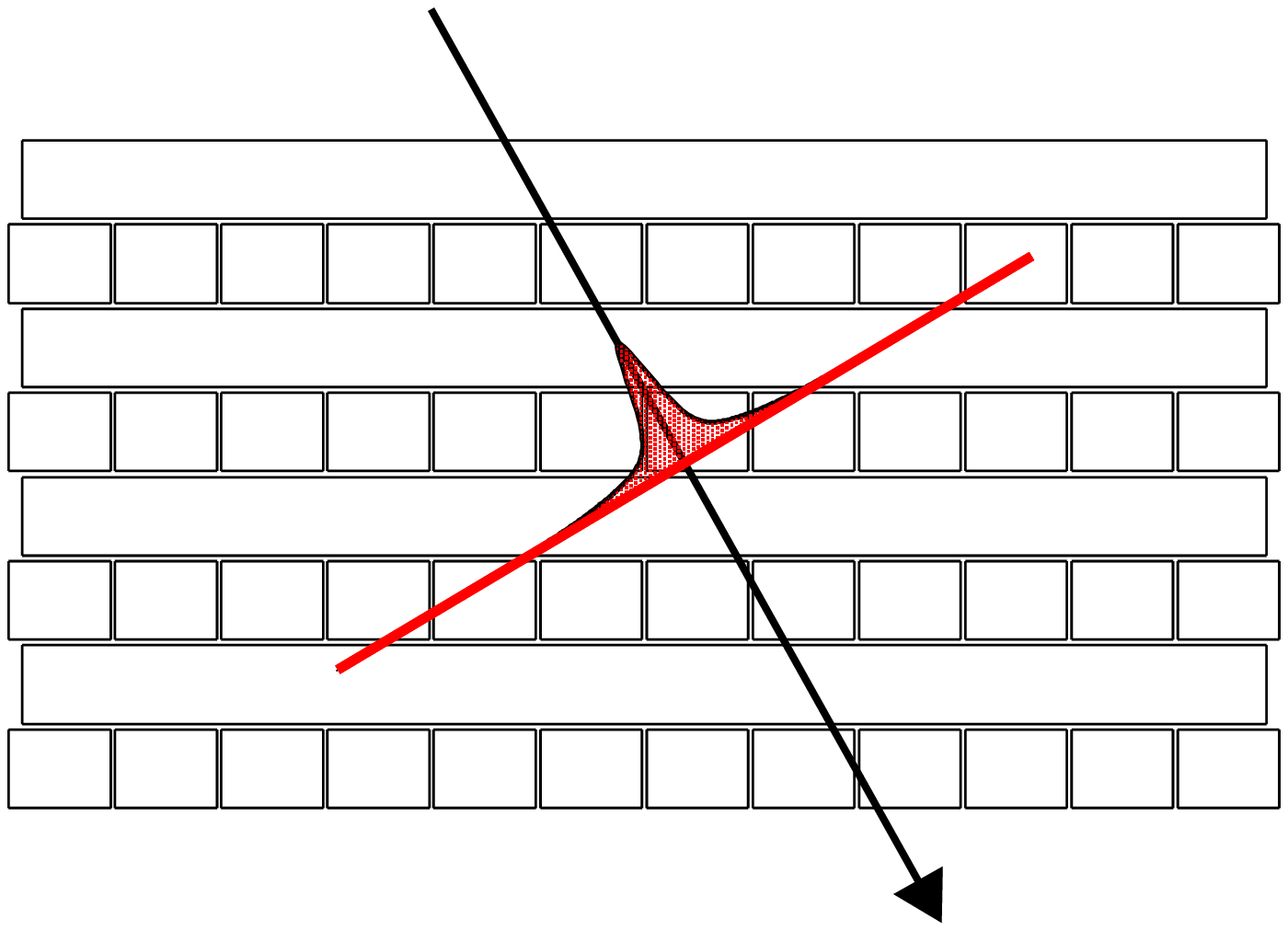}
\caption{\label{fig:devthroughcal}One step of the computation of the development of an individual shower inside a calorimeter module. The arrow corresponds to the photon trajectory. The red disk represents a shower slice at a  given shower depth for which we compute the fraction of energy deposited in each layer.}
\end{minipage}\hspace{2pc}%
\begin{minipage}{20pc}
\includegraphics[width=20pc]{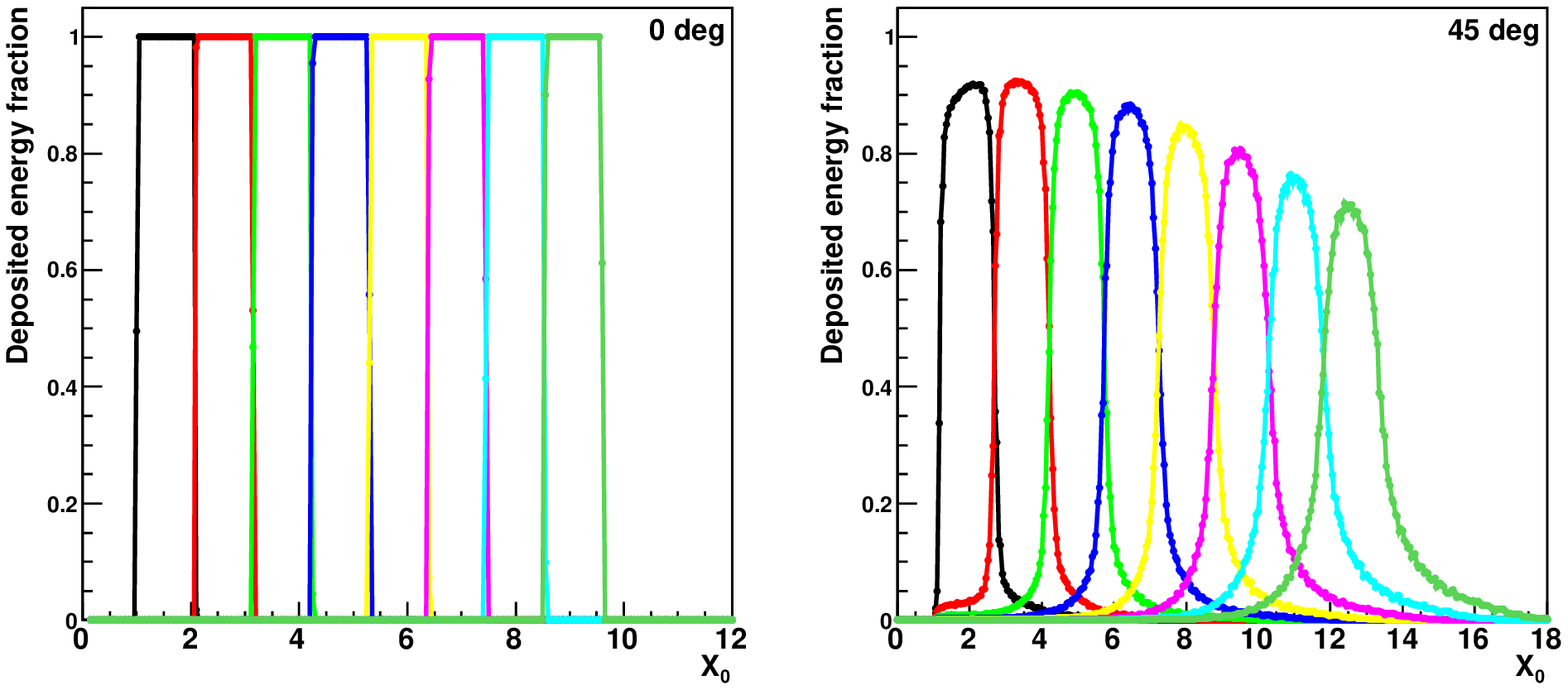}
\caption{\label{fig:afterconvertfixedx0}Predicted fraction of energy deposited in the layers of the calorimeter for a photon at $0^\mathrm{o}$ (left) and $45^\mathrm{o}$ (right) as a function of the shower depth. Each curve corresponds to one layer: from the first layer on the left to the last layer on the right.}
\end{minipage} 
\end{figure}

Using this trajectory information, the computation of $f_i(t)$ is done as follows (illustrated in Fig.~\ref{fig:devthroughcal}):
\begin{itemize}
\item divide the trajectory into $\sim$~1.85~mm steps, starting 100~cm ahead of the entry point of the photon in the calorimeter;
\item at each step along the trajectory, at a distance $x$ in mm along the trajectory from the starting point of the shower, corresponding to the shower depth $t$:
  \begin{itemize}
  \item define a disk centered on the step point, perpendicular to the trajectory, whose radius is 3 Moliere radii = 95~mm (the red line in Fig.~\ref{fig:devthroughcal});
  \item define a grid of test points on the disk, with a step of 0.05 Moliere radius;
  \item the test points are used to compute the fraction of energy deposited in the layers at the shower depth $t$, using the transverse profile parameterization depending on $t/T$ (represented in Fig.~\ref{fig:devthroughcal} by the red area); 
  \item compute $\mathrm dt$, the number of radiation lengths corresponding to the longitudinal step, taking into account the active (CsI cystals) and passive (gaps between calorimeter modules) material seen by the shower, and add it to $t$.
  \end{itemize}
\end{itemize}

While developing through the calorimeter, the shower becomes wider. At some point, two different parts of the shower at the same shower depth $t$ evolve independently. In other words, if one part of the shower escapes the calorimeter, the rest of the shower evolves as if the whole shower was inside the calorimeter. This is taken into account when computing $\mathrm dt$.  At this stage of the computation, we have the energy fractions as a function of $x$ as well as the shower depth $t$ as a function of $x$. We then deduce the energy fractions as a function of $t$: $f_i(t)$.

Examples of $f_i(t)$, the fractions of energy deposited in the layers, are shown in Fig.~\ref{fig:afterconvertfixedx0}, for a shower maximum $T$ equal to 8~radiation lengths. The fraction of energy deposited in layer $i$ at normal incidence is very simple, being 1 when the shower is inside layer $i$, and 0 otherwise. For an incidence angle of 45~degrees, the situation is more complex, because of the sharing of the energy between layers, which increases with $t$ as the shower becomes wider.

\subsection{Crystal saturation handling}
\label{section:saturation}

Due to the readout limitation, the signal of a crystal saturates when the deposited energy is greater than $\sim70$~GeV. It occurs for photons with $E>1$~TeV. In order to take this saturation effect into account in the shower fit, we consider the saturated crystals separately, and especially take out their energy from the layer energies. The computation of the shower development through the calorimeter (presented in section~\ref{section:devthroughcal}) is used to determine the fraction of energy deposited in the saturated crystals as a function of $t$. The standard part of the $\chi^2$ becomes:
$$ \chi^2 = \sum_\mathrm{layers} \left(e_{m,i}-e_{p,i} \right)^2/\delta e^2 + \sum_\mathrm{sat. crystals} \left(\mathrm{max}(0,e_{m,i}-e_{p,i}) \right)^2/\delta e^2 $$
where the layer energies are summed over the non saturated crystals and where we introduced the term $\mathrm{max}(0,e_{m,i}-e_{p,i})$ in order that the saturated crystal contribution vanishes as soon as the predicted energy is greater than the measured energy.

\subsection{Results}

\begin{figure}[h]
  \begin{center}
    \includegraphics[width=0.90\textwidth]{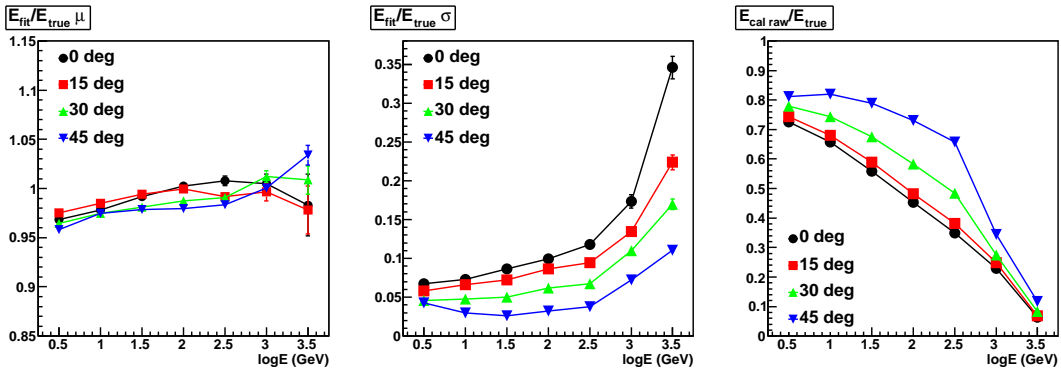}
    \caption{Performance of the shower profile fit: bias (left), resolution (center), shower containment (right).}
    \label{fig:getresolution1}
  \end{center}
\end{figure}

The performance of the shower profile fit is summarized in Fig~\ref{fig:getresolution1}. The bias is less than 3\% and is easily corrected for. The energy resolution varies with energy and incidence angle. At normal incidence, it degrades when the energy increases. This is because the shower containment decreases with energy, as shown in Fig~\ref{fig:getresolution1}(right). At $45^\mathrm{o}$, the resolution improves slightly from 3~GeV to 30~GeV because the fraction of energy deposited in the tracker decreases. Above 30~GeV, the resolution starts to degrade because the shower containement decreases. It must be noted that at 3~TeV, the shower containement is only $\sim10$\% and the mean number of saturated crystals is $\sim6$. But still we have a decent energy resolution, especially for off-axis photons.

\section{Some Fermi results at high energy}
It is impossible to summarize all the results obtained by Fermi after 4 years of operation. We only focus on results at high energy for which the measurement of the energy is important.

\subsection{In-flight absolute energy scale}
There is no gamma-ray source with a sharp and well known spectral feature for absolute energy scale calibration. But we can use the cutoff of the electron spectrum due to the Earth's magnetic field to check our absolute energy scale~\cite{2012APh....35..346A}. As explained in section~\ref{section:electronspectrum}, Fermi can detect electrons and positrons as efficiently as gamma rays. The geomagnetic cutoff rigidity depends on the geographic longitude and latitude. The altitude of the Fermi orbit is 565~km and its inclination is $26.5^\mathrm{o}$. So the Fermi spacecraft spans a range of geomagnetic cutoff that goes from 6 to 14~GeV. This range is divided into 6 bins. In each geomagnetic cutoff bin, we first determine the expected primary electron+positron spectrum. This is done by using a tracer code that computes the trajectories in the reverse direction (from the spacecracft to space) in order to reject secondary particles that are produced in the atmosphere. The measured primary spectrum is obtained after subtraction of the secondary fraction which is derived using the azimuthal distribution. The ratio of the cutoffs of the expected and measured spectra is shown in Fig.~\ref{fig:poseleexp}. From this comparison, we find that, in the energy range [6,14~GeV], the energy scale is known within 5\%.

\subsection{Gamma-ray sources spectral features}
By detecting more than 1800 sources, Fermi is able to peform detailed population studies. Concerning active galactic nuclei, the most numerous sources for Fermi, spectral breaks have been observed to be common in Flat Spectrum Radio Quasars (the break energy ranging from 1 to 10 GeV in the source frame for the brightest sources), and present also in some bright low-synchroton peaked BL~Lacs~\cite{2011ApJ...743..171A}. These breaks represent challenges for theoretical models aiming at describing the blazar phenomenon. Another important class of objects are pulsars: Fermi has currently detected more than 100 pulsars, whereas EGRET only detected 6. Pulsars have a power-law spectrum with a cutoff, which position and shape depends on the location of the gamma-ray emission region in the pulsar magnetosphere. The cutoffs of all pulsars measured by Fermi~\cite{2010ApJS..187..460A} are consistent with exponential cutoffs at a couple of GeV. This result rules out the low-altitude emission models which predict a sharper cutoffs.

\subsection{Gamma-ray bursts and Lorentz invariance violation}
Quantum gravity may cause the violation of Lorentz invariance and the speed of light could then depend on energy. It would imply that photons with different energies emitted simultaneously by a very distant source would not arrive at the same time. Fermi is able to constrain the violation of Lorentz invariance by looking at the highest energy events of gamma-ray bursts. GRB090510~\cite{2009Natur.462..331A} was a short gamma-ray burst with a redshift $z = 0.9$. The highest energy event has 31~GeV and is detected 0.86~ms after the start of the emission below 1~MeV (measured by the Gamma-ray Burst Monitor, the second instrument on-board Fermi). In the framework of linear Lorentz invariance violation, the delay between two photons is proportional to the energy difference divided by $M_\mathrm{QG}$, the quantum gravity mass scale. Assuming that the high energy emission did not start before the burst, this 31~GeV event sets the limit: $M_\mathrm{QG}/M_{Planck}>1.2$.

\subsection{Earth limb spectrum}
Gamma rays from the Earth limb are the result of the interaction of cosmic rays in the atmosphere. Above 10~GeV, the emission that is detected by Fermi is only due to grazing incidence cosmic rays in the upper atmosphere. The altitude range of the emission is rather thin: cosmic-rays passing too high do not interact enough and the gamma-rays produced by cosmic-rays that pass too low are attenuated by interacting in the atmosphere. The measured Earth limb spectrum~\cite{2009PhRvD..80l2004A} is a power law with spectral index = $2.79 \pm 0.06$, in agreement with the spectral index of cosmic-ray proton = 2.75. The spectrum is very smooth up to 1 TeV. The absence of any significant sharp feature is an additional proof that our photon selection is well under control.

\subsection{Extragalactic diffuse emission}
The largest source of Fermi photons is the galactic diffuse emission, due to the interaction of cosmic-rays with the matter in the Galaxy. But there is a much fainter diffuse emission that has an extragalactic origin and that represents the history of the non-thermal universe. Thanks to a careful background subtraction, we measured the extragalactic diffuse emission up to 100~GeV~\cite{2010PhRvL.104j1101A} and expanding this measurement up to 1~TeV is under way. Currently, unresolved sources like blazars, radio galaxies and star-forming regions account for 50\% to 80\% of this emission. 

\subsection{Cosmic-ray electron and positron spectrum}
\label{section:electronspectrum}
Since Fermi's primary goal is to detect gamma-rays, the main trigger configuration rejects a large fraction of charged particles by requiring that there is no activity in the ACD above the tracker tower that triggered. But, in order to be fully efficient at high energy, all events that deposit more than 20~GeV in the calorimeter are sent to the ground, which is the case for high energy electrons and positrons. Since they produce electromagnetic showers very similar to photon induced showers, Fermi is able to reconstruct and select electrons and positrons a well as gamma-rays. The fraction of residual background is estimated to vary smoothly from 4\% at 20~GeV to 20\% at 1~TeV. Since measuring such high energies with a 8.6~radiation length thick calorimeter is not an easy task, we checked that our results did not change when selecting off-axis events that pass through more than 12~radiation lengths in the calorimeter. The measured spectrum~\cite{2010PhRvD..82i2004A} rules out the excess around 700~GeV seen by the ATIC experiment~\cite{ATIC}
 and can be explained by a conventional model with an adjusted injection spectrum or with an additional source, which could be a nearby pulsar.

\begin{figure}[h]
\begin{minipage}{18pc}
\includegraphics[width=18pc]{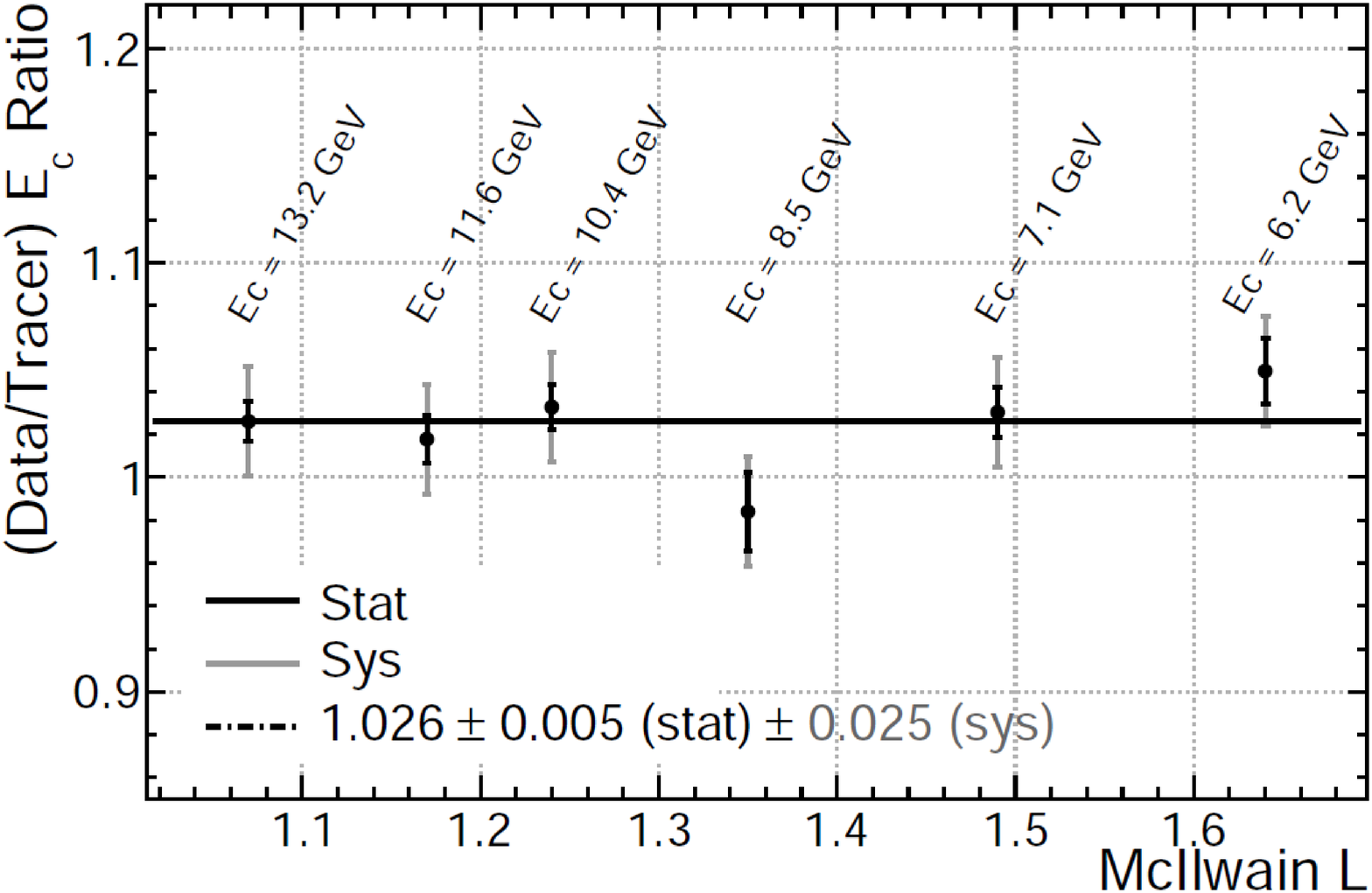}
\caption{\label{fig:absenescale}Distribution of cutoff energy ratios as a function of McIlwain L (with corresponding cutoff energies indicated). The statistical errors are represented by the black error bars, and the systematic errors are in gray. From~\cite{2012APh....35..346A}}
\end{minipage}\hspace{2pc}%
\begin{minipage}{18pc}
\includegraphics[width=18pc]{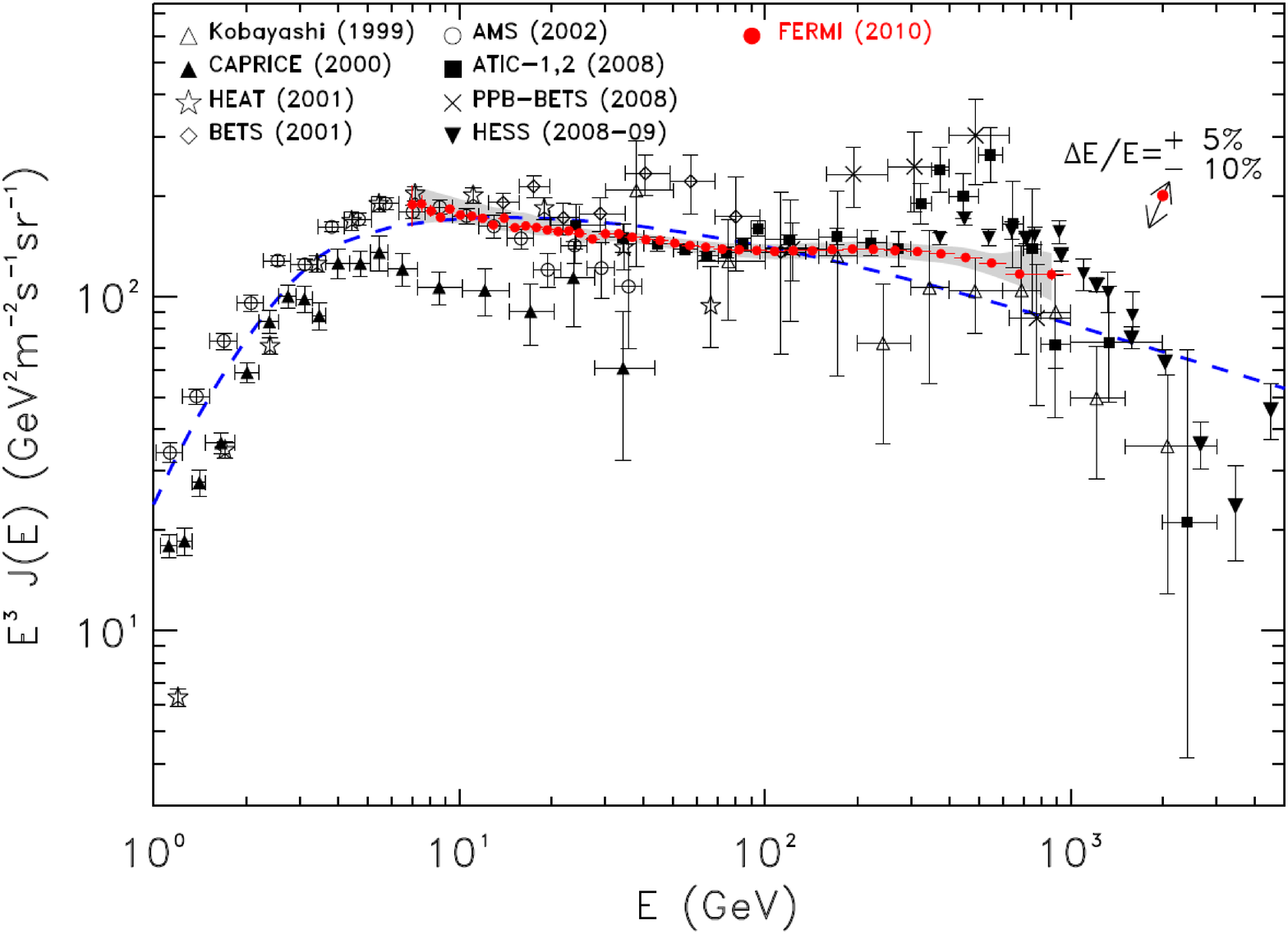}
\caption{\label{fig:fullelec.eps}Cosmic-ray electron spectrum as measured by Fermi LAT for 1 yr of observations along with other recent high-energy results. Systematic errors are shown by the gray band. From~\cite{2010PhRvD..82i2004A}.}
\end{minipage} 
\end{figure}

\subsection{Cosmic-ray positron fraction}
\begin{figure}[h]
\begin{minipage}{18pc}
\includegraphics[width=18pc]{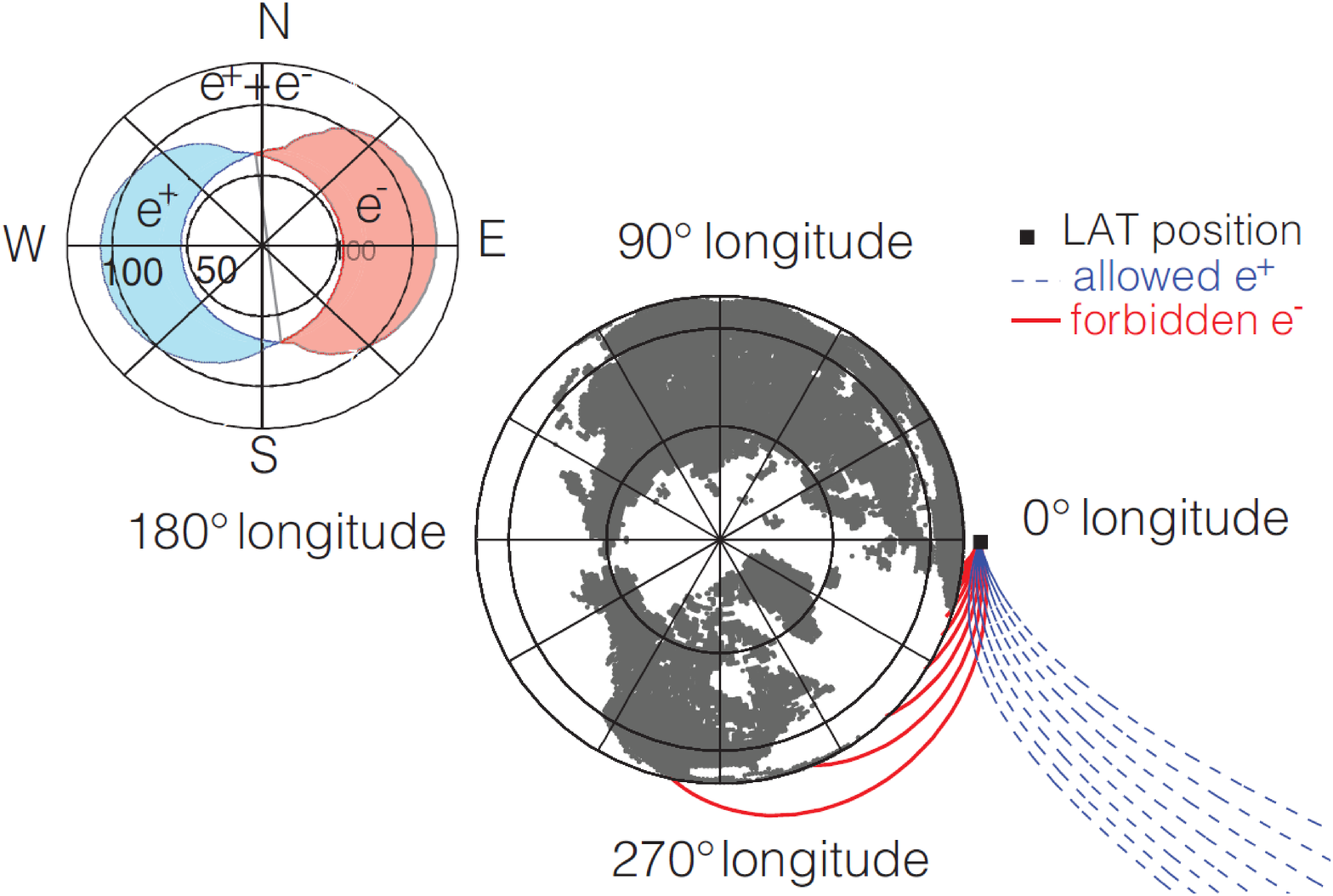}
\caption{\label{fig:poseleexp}Examples of calculated electron (red) and positron (blue) trajectories arriving at the detector, for 28 GeV particles arriving within the Equatorial plane (viewed from the North pole). Forbidden trajectories are solid and allowed trajectories are dashed. From~\cite{2012PhRvL.108a1103A}.}
\end{minipage}\hspace{2pc}%
\begin{minipage}{18pc}
\includegraphics[width=18pc]{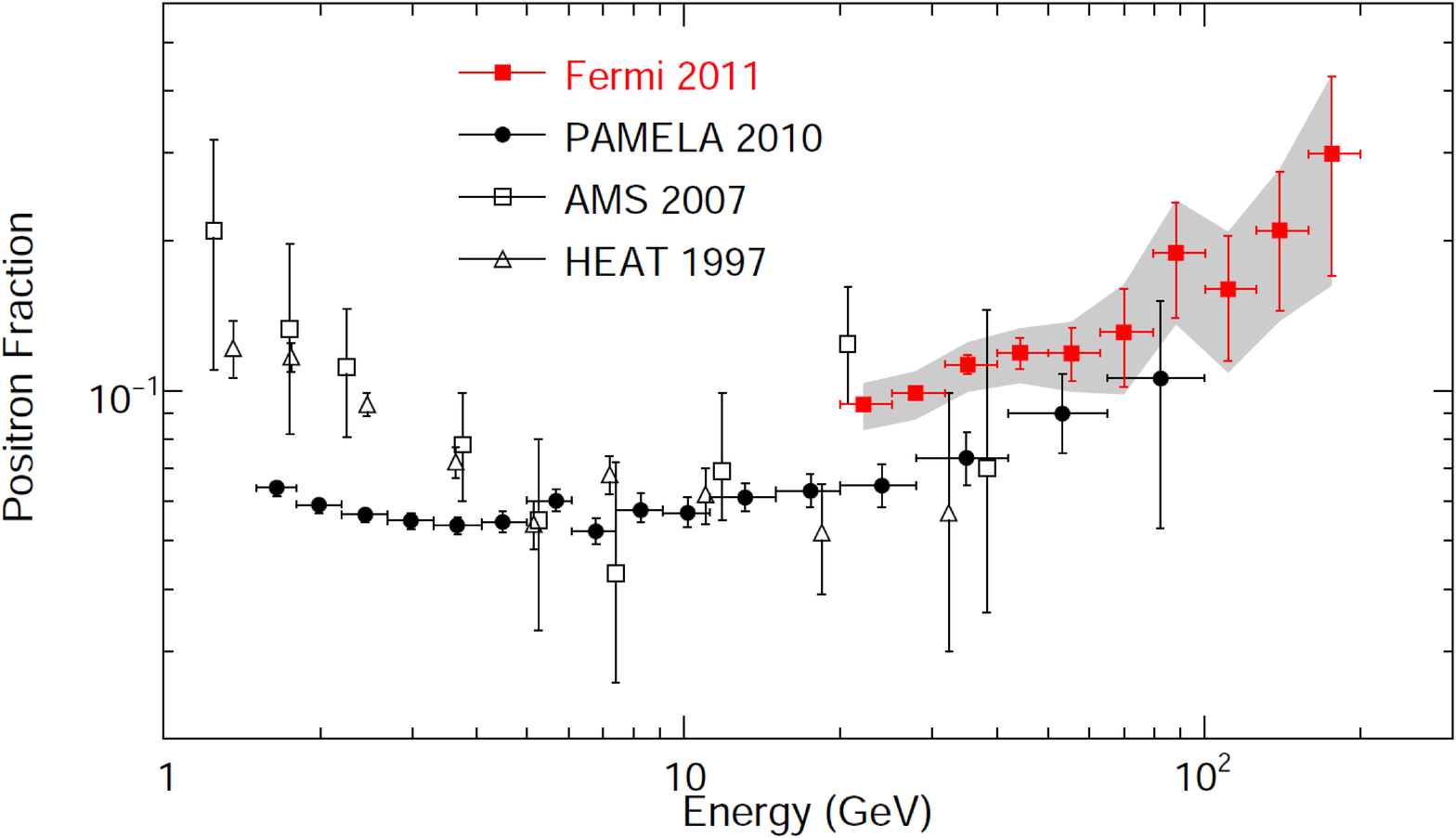}
\caption{\label{fig:poseleratio}Positron fraction measured by the Fermi LAT and by other experiments. The Fermi statistical uncertainty is shown with error bars and the total (statistical plus systematic uncertainty) is shown as a shaded band. From~\cite{2012PhRvL.108a1103A}. }
\end{minipage} 
\end{figure}

Without any magnet on-board, Fermi is not able to discriminate between electrons and positrons. But because of the geomagnetic field, there is a region of the Earth's horizon from which only electrons can come and another region from which only positrons can come, as shown in Fig.~\ref{fig:poseleexp}. So Fermi is able to measure separately the spectra of electrons and positrons between 20~GeV and 200~GeV (above 200~GeV these regions become too small). From these spectra we can deduce the positron fraction, which is shown in Fig.~\ref{fig:poseleratio}. Our measurement~\cite{2012PhRvL.108a1103A}, which was the first above 100~GeV, confirms the increase of the fraction with energy as reported by PAMELA~\cite{PAMELA}.

\section{Conclusion}
After 4 years in orbit, Fermi has proved to be an excellent gamma-ray telescope. Thanks to its large effective area and field of view, to its good angular and energy resolution, Fermi has already unravelled the gamma-ray sky with much more precision than its predecessor EGRET. By detecting more than 1800 sources, it allows population studies of pulsars, supernova remnants, active galactic nuclei and gamma-ray bursts. The diffuse emission of the Galaxy and the extragalctic diffuse emission are also known much better. Fermi has proved to be a very good detector of electrons and positrons as well, and, by using the geomagnetic field, it is even able to discriminate between them. Part of the success of Fermi is due to its large energy range. Thanks to the fine segmentation of its calorimeter, we are able to measure and select gamma rays, electrons and positrons up to 3~TeV, despite the modest thickness of the calorimeter. Taking benefit from our experience after 4 years in orbit, we are currently optimizing the event reconstruction and selection. With a better sensitivity, Fermi will certainly continue to improve our understanding of the high energy sky.

\section*{References}
\bibliography{calor2012_proceeding}

\end{document}